\documentclass[aps,prl,twocolumn,showpacs,floatfix,superscriptaddress,notitlepage]{revtex4-2}
\usepackage[utf8]{inputenc}
\usepackage{soul}
\usepackage[left=2cm,right=2cm]{geometry}
\usepackage{physics}
\usepackage{amsmath}
\usepackage{amsfonts}
\usepackage{amssymb}
\usepackage{mathtools}
\usepackage{xcolor}
\usepackage{graphicx}
\usepackage[caption=false]{subfig}
\usepackage{dsfont}
\usepackage{amsthm}
\usepackage[hidelinks]{hyperref}
\usepackage{textcomp}
\usepackage{url}
\usepackage[normalem]{ulem}
\usepackage{booktabs}

\usepackage{siunitx}
\DeclareSIUnit\Gal{Gal}

\hypersetup{%
  colorlinks=true,
  linkcolor=blue,
  citecolor=blue,
  urlcolor=blue,
}

\makeatletter
\def\@fnsymbol#1{\ensuremath{\ifcase#1\or
  \dagger\or
  \ddagger\or
  \mathsection\or
  \mathparagraph\else
  **\fi}}
\makeatother

\begin{document}

\title{Field-Deployable Hybrid Gravimetry: Projecting Absolute Accuracy Across a Remote 24km$^2$ Survey via Daily Quantum Calibration}

\author{Nathan Shettell}
\thanks{contact: nathans@nus.edu.sg}
%\thanks{These authors contributed equally to this work.}
\affiliation{Centre  for  Quantum  Technologies,  National  University  of  Singapore,  Singapore 117543, Singapore}
%\email{} %incase we want someone to be corresponding author

\author{Kai Sheng Lee}
%\thanks{These authors contributed equally to this work.}
\affiliation{Centre  for  Quantum  Technologies,  National  University  of  Singapore,  Singapore 117543, Singapore}

\author{Fong En Oon}
\affiliation{Centre  for  Quantum  Technologies,  National  University  of  Singapore,  Singapore 117543, Singapore}

\author{Elizaveta Maksimova}
\affiliation{Centre  for  Quantum  Technologies,  National  University  of  Singapore,  Singapore 117543, Singapore}

\author{Hong Hui Chen}
\affiliation{Centre  for  Quantum  Technologies,  National  University  of  Singapore,  Singapore 117543, Singapore}

\author{Rainer Dumke}
\affiliation{Centre  for  Quantum  Technologies,  National  University  of  Singapore,  Singapore 117543, Singapore}
\affiliation{School  of  Physical  and  Mathematical  Sciences,  Nanyang  Technological  University,  637371, Singapore}

%\date{\today}

\begin{abstract}
Hybrid gravimetry combines the mobility of relative gravimeters with the long-term stability of an absolute reference. Following our previous demonstration of this concept under controlled field conditions~\cite{Shettell2024}, we investigate whether quantum-assisted hybrid gravimetry retains its calibration performance during an extended survey under realistic operational constraints. A transportable atomic gravimeter provided routine \si{\micro\Gal}-level calibration of two mobile spring gravimeters throughout a seven-day survey spanning 24 km$^2$ of dense tropical terrain. Despite rainfall, dense canopy, and limited field infrastructure, the atomic reference enabled high-precision asynchronous comparison of relative measurements while suppressing instrumental drift in regions where GPS positioning remained reliable. These results demonstrate that transportable atomic gravimeters can provide practical calibration references for extended geophysical surveys, extending hybrid quantum gravimetry beyond controlled demonstrations toward realistic field deployment.
\end{abstract}

\maketitle

% \begingroup
% \renewcommand\thefootnote{\dagger}
% \footnotetext{These authors contributed equally to this work.}
% \endgroup

\section{Introduction}
Gravimetry provides a passive, non-invasive means of probing subsurface density variations and plays a central role in applications including geological mapping, groundwater monitoring, and crustal deformation studies \cite{hinze2013gravity, van2017geophysics, flechtner2021satellite, davis2008time, timmen2012observing}. Field surveys are typically performed using either relative or absolute gravimeters \cite{Niebauer2015}. Relative gravimeters are compact and well suited to dense spatial sampling but suffer from time-dependent instrumental drift, whereas absolute gravimeters provide intrinsically drift-free measurements of gravitational acceleration \cite{Niebauer2015, lacoste1988zero, gillot2014stability}. Among the latter, atomic gravimeters exploit matter-wave interferometry to achieve high precision and long-term stability, although their complexity and deployment requirements have limited their routine use in large-area field surveys \cite{kasevich1991atomic, peters1999measurement, menoret2018gravity, ruan2024transportable, Shettell2024, chen2025joint}.

Hybrid gravimetry has recently emerged as a promising strategy for combining the mobility of relative gravimeters with the long-term stability of an absolute reference \cite{Shettell2024, sugihara2008geothermal, nishijima2016repeat, greco2012combining, zerbini2002multi, francis1998calibration, imanishi2002calibration, niebauer2011simultaneous, merlet2021calibration, chen2025joint}. In Ref.~\cite{Shettell2024}, we demonstrated continuous calibration of spring gravimeters using an on-site atomic gravimeter during a two-day urban field campaign, establishing the feasibility of quantum-assisted hybrid gravimetry under controlled conditions. Practical gravity surveys, however, are typically conducted over longer durations and in far more demanding environments, where adverse weather, dense vegetation, limited infrastructure, and degraded positioning can compromise measurement quality. Here, we investigate whether hybrid gravimetry retains its calibration performance under these realistic operating conditions.

We report a hybrid gravity survey spanning 24 km$^2$ of dense tropical terrain over multiple days. A compact atomic gravimeter operated continuously as an absolute reference while two CG6 spring gravimeters \cite{hugill1990scintrex} performed asynchronous measurements throughout the campaign. Despite rainfall, dense canopy, and limited field infrastructure, periodic calibration against the atomic gravimeter enabled microGal-level drift correction in regions where GPS positioning remained reliable. These results demonstrate that transportable quantum gravimeters can provide practical calibration references for extended geophysical surveys beyond controlled field demonstrations.

\section{Methods}

Prior to data collection, a site reconnaissance was conducted to identify a suitable location for the atomic gravimeter base station. Selection criteria were driven by both operational and metrological requirements: sufficient space to house an air-conditioned cargo container for the quantum gravimeter (75 cm $\times$ 75 cm $\times$ 200 cm), provision for continuous power via a generator, and mechanically stable, level ground to minimize tilt-induced bias over the duration of the deployment. The survey was confined to a 24 km$^2$ area of remote, forested terrain on an island off the coast of Singapore; the exact geographic coordinates are not disclosed.

\begin{figure}[t!]
    \centering
    \includegraphics[height=4.5cm]{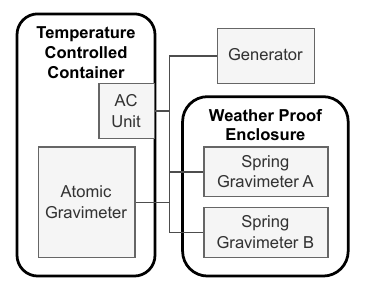}
\caption{Layout of the reference station used for daily calibration. The atomic gravimeter was housed inside an air-conditioned container (approximately 75 cm $\times$ 75 cm $\times$ 200 cm) and operated continuously to provide an absolute gravity reference, with the container positioned on a cemented clearing to ensure stability. Spring gravimeters were placed in a neighbouring weather-protected, animal-resistant metal enclosure during overnight calibration periods, enabling correction for instrumental drift. A common generator provided electrical power throughout the survey.}
\label{fig:layout}
\end{figure}

The initial phase of the campaign (2–5 October 2023) was dedicated to installation, alignment, and stabilization of the atomic gravimeter. Field gravity measurements were subsequently conducted between 6 and 12 October 2023, with daily survey operations carried out between 10:00 and 17:00, weather permitting. Two CG6 spring gravimeters were used for mobile measurements and returned each evening to a fixed location adjacent to the atomic gravimeter base station. During overnight periods (approximately 18:00 to 09:00), both the atomic and spring gravimeters remained stationary and recorded continuous gravity data at one-minute intervals. Because all instruments sampled the same local gravitational field under stable conditions, these measurements provided a common reference for monitoring and correcting time-dependent drift in the relative gravimeters \cite{schilling2015accuracy, Shettell2024}. This calibration scheme enabled high-precision cross-comparison between mobile measurements acquired on different days. The instrument configuration at the reference station is depicted in Fig.~(\ref{fig:layout}).

% Initial GPS trials revealed significant signal degradation under dense canopy cover, a consequence of the survey area's remoteness and limited infrastructure. Because accurate positioning required a clear line of sight to overhead satellites, measurement points were restricted to exposed roadways and walking paths with minimal foliage overhead.

\subsection{Measurements} 
Two spring gravimeters were operated concurrently to measure local variations in gravitational acceleration along separate survey paths. Measurement stations were spaced at least 50~m apart along predefined survey axes. At each station, the gravimeter was leveled by adjusting its tripod until the internal tilt readings were within $\pm$10 arcseconds, after which the instrument was observed for approximately 30 s to verify mechanical stability. This stability check was necessary under the tropical field conditions, where rainfall and humid, unconsolidated topsoil could cause slight tripod settling during measurements. If significant drift was detected, the instrument was repositioned to a more stable surface and re-leveled. Once a stable setup was achieved, gravity data were acquired over a 5 min interval, with five measurements recorded at one-minute cadence.

\subsection{GPS}
Precise latitude, longitude, and elevation information were required to resolve spatial variations in the gravity field and to compute terrain-related corrections, including Bouguer anomalies \cite{chapin1996theory}. Dense canopy cover across much of the survey area limited satellite visibility, necessitating GPS measurements at locations with sufficient sky exposure. Each spring gravimeter was fitted with a rigid metal bracket supporting a GPS antenna, enabling position data to be recorded concurrently with each gravity measurement, while a continuously operating GPS receiver on a nearby building served as a fixed reference station for post-processed kinematic (PPK) analysis \cite{pillai2022potential}.

PPK processing combined reference-station and mobile-antenna data to refine the position estimates for each measurement. Solutions either achieved centimeter-level vertical precision, provided reduced or unquantified vertical accuracy, or failed to converge; only the latter were excluded from further analysis. Stations with centimeter-level precision provided high-quality elevation control for validating the atomic gravimeter as a stable temporal reference, while stations with reduced accuracy were retained to preserve spatial coverage and support analysis of longer-wavelength gravity variations across the survey area.

\section{Data Correction}

Understanding gravity variations across the survey area requires careful correction for both temporal and spatial (elevation and terrain dependent) effects, ensuring that the residual gravity signal can be confidently attributed to subsurface mass distributions. In gravimetry, only horizontal variations are typically of interest, while vertical variations, instrumental drift, and tidal effects must be removed. Additionally, spring gravimeters measure relative gravity rather than absolute values, necessitating calibration against an absolute reference to ensure consistency across instruments and survey periods.

\begin{table*}[t]
    \centering
    \small
    \setlength{\tabcolsep}{6pt}
    \renewcommand{\arraystretch}{1.15}
    \begin{tabular}{l c l}
        \toprule
        Source & Magnitude & Description \\
        \midrule
        Baseline error          & $3$--$8~\si{\micro\Gal}$   & Observed effects of deployment and field conditions \\
        Elevation correction    & $\leq 25~\si{\micro\Gal}$  & GPS elevation uncertainty $\leq 10$ cm \\
        Terrain correction      & $\lesssim 1~\si{\micro\Gal}$ & Estimated from low-relief terrain \\
        Spring drift correction & $\leq 5~\si{\micro\Gal}$   & Residual after daily calibration \\
        Atomic reference        & $\sim 4~\si{\micro\Gal}$   & Field Allan deviation \\
        Tidal correction        & $<1~\si{\micro\Gal}$       & Model uncertainty \\
        \bottomrule
    \end{tabular}
    \caption{Summary of the principal uncertainty contributions to the hybrid gravity measurements.}
    \label{tab:uncertainty_budget}
\end{table*}

% \begin{table}[t]
%     \centering
%     \small
%     \setlength{\tabcolsep}{3pt}
%     \renewcommand{\arraystretch}{1.15}
%     \begin{tabular}{l c p{0.38\columnwidth}}
%         \toprule
%         Source & Magnitude & Description \\
%         \midrule
%         Baseline Error 
%             & $3$--$8~\si{\micro\Gal}$ 
%             & Observed effects of deployment and field conditions \\
%         Elevation correction 
%             & $\leq 25~\si{\micro\Gal}$ 
%             & GPS elevation uncertainty $\leq 10$ cm \\
%         Terrain correction 
%             & $\lesssim 1~\si{\micro\Gal}$ 
%             & Estimated from low-relief terrain \\
%         Spring drift correction 
%             & $\leq 5~\si{\micro\Gal}$ 
%             & Residual after daily calibration \\
%         Atomic reference 
%             & $\sim 4~\si{\micro\Gal}$ 
%             & Field Allan deviation \\
%         Tidal correction 
%             & $<1~\si{\micro\Gal}$ 
%             & Model uncertainty \\
%         \bottomrule
%     \end{tabular}
%     \caption{Summary of the principal uncertainty contributions to the hybrid gravity measurements.}
%     \label{tab:uncertainty_budget}
% \end{table}

Without loss of generality, the spring gravimeter measurements at horizontal position $\vec{x}$ (latitude and longitude) and time $t$ can be expressed as
\begin{equation}
    g_\text{sg}(\vec{x},t) = g_0(\vec{x})+\delta_t g(\vec{x},t)+\delta_z g(\vec{x})+\Delta_\text{sg}(t),
\end{equation}
where
\begin{enumerate}
    \item $g_0(\vec{x})$ is the absolute gravity at position $\vec{x}$, which represents the desired signal after all corrections;
    \item $\delta_t g(\vec{x},t)$ captures time-dependent variations in the gravity field at position $\vec{x}$ arising from geophysical and environmental processes, e.g., tidal dynamics and ocean loading;
    \item $\delta_z g(x)$ accounts for elevation-dependent spatial corrections, such as free-air and Bouguer corrections applied to reference measurements;
    \item $\Delta_\text{sg} (t)$ represents instrument-specific, slowly varying temporal effects associated with the spring gravimeter, including bias offsets and long-term drift. Over short timescales this term is typically dominated by an approximately linear drift, and it is continuously constrained through calibration against the atomic gravimeter to ensure faithful transfer of long-term gravity variations into the relative data.
\end{enumerate}
In this analysis, all gravity measurements are referenced to mean sea level, as determined from GPS-derived elevations. Data correction is therefore divided into two components: i) \emph{Temporal corrections}, addressing instrumental drift, tidal effects, and inter-instrument calibration; and ii) \emph{Elevation corrections}, accounting for vertical variations in gravitational acceleration due to changes in elevation. The field measurements have a baseline uncertainty of approximately $3$-$8~\si{\micro\Gal}$, primarily reflecting measurement and deployment effects such as settling of the instruments in the topsoil. The corrections introduce additional uncertainty contributions, summarized in Table~\ref{tab:uncertainty_budget} and discussed below, which are combined with the baseline uncertainty through standard error propagation.

\subsection{Temporal Correction}

% \begin{figure*}
%     \centering
%     \includesvg[width=0.95\textwidth]{DriftCorrection.svg}
% \caption{Overnight calibration data for the two spring gravimeters, $g_\text{sgA}$ (a) and $g_\text{sgB}$ (b), compared against continuous absolute gravity measurements from the atomic gravimeter $g_\text{ag}$ (c), collected over eight days. Each spring gravimeter panel shows uncorrected data (red), a single global drift correction from the start of the survey (blue), and daily drift corrections derived from the atomic reference (green). All data are tide-corrected and plotted as relative gravity to emphasize temporal variations. The daily calibration accurately tracks the gradual \SI{50}{\micro\Gal} increase observed by the atomic gravimeter, while the single global correction does not. A brief discontinuity in $g_\text{sgB}$ near the end of the survey resulted from a power interruption and was corrected during processing. Panel (d) shows the Allan deviation of the atomic gravimeter measurements in the field (circles) compared with laboratory data over a similar timescale (squares), demonstrating comparable long-term stability with slightly elevated white-noise levels in the field due to environmental vibrations and tilt.}
% \label{fig:driftcorrection}
% \end{figure*}

\begin{figure*}
    \centering
    \includegraphics[width=0.95\textwidth]{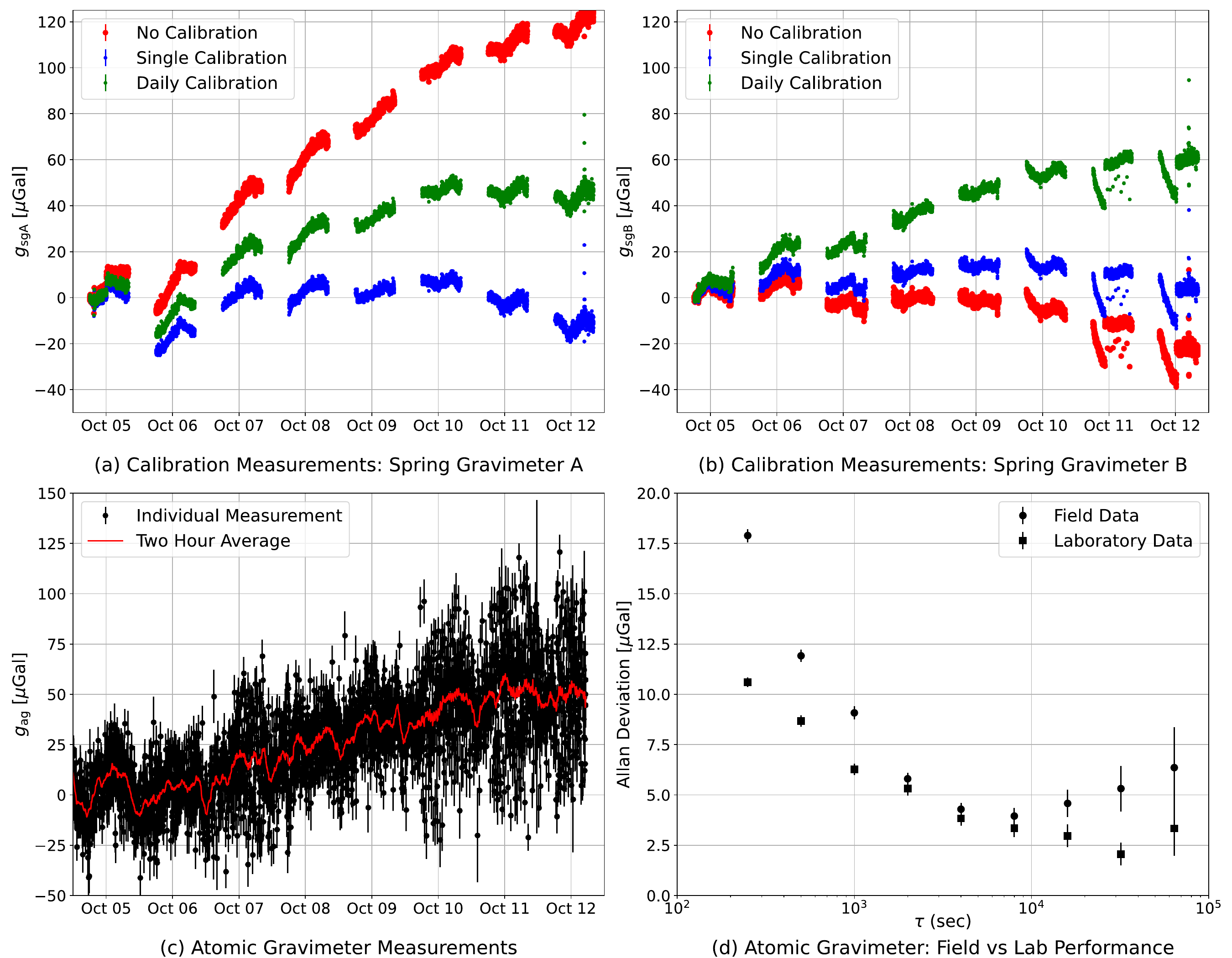}
\caption{Overnight calibration data for the two spring gravimeters, $g_\text{sgA}$ (a) and $g_\text{sgB}$ (b), compared against continuous absolute gravity measurements from the atomic gravimeter $g_\text{ag}$ (c), collected over eight days. Each spring gravimeter panel shows uncorrected data (red), a single global drift correction from the start of the survey (blue), and daily drift corrections derived from the atomic reference (green). All data are tide-corrected and plotted as relative gravity to emphasize temporal variations. The daily calibration accurately tracks the gradual \SI{50}{\micro\Gal} increase observed by the atomic gravimeter, while the single global correction does not. The dominant instrumental drift of spring gravimeters over these calibration intervals is well described by a linear trend \cite{hugill1990scintrex, gabalda2003cg3tool}; residual deviations are therefore incorporated into the uncertainty of the drift correction rather than modeled using higher-order terms. A brief discontinuity in $g_\text{sgB}$ near the end of the survey resulted from a power interruption and was corrected during processing. Panel (d) shows the Allan deviation of the atomic gravimeter measurements in the field (circles) compared with laboratory data over a similar timescale (squares), demonstrating comparable long-term stability with slightly elevated white-noise levels in the field due to environmental vibrations and tilt.}
\label{fig:driftcorrection}
\end{figure*}

Two primary sources of temporal variation in gravity measurements are Earth tides \cite{mikolaj2013first, jentzsch2005earth} and instrumental drift in the spring gravimeters \cite{schilling2015accuracy}. Tidal effects, which can reach amplitudes up to \SI{150}{\micro\Gal}, were removed from both absolute and relative gravity data using the external model of Micro-g LaCoste’ QuickTide Pro.

Instrumental drift was corrected using the continuously recording atomic gravimeter as an on-site absolute reference. Each evening, the spring gravimeters were co-located with the atomic instrument and left recording overnight under stable conditions (18:00–07:00), enabling direct comparison and quantification of instrument-specific drift. The observed drift over these short calibration intervals was predominantly linear and was therefore modeled as
\begin{equation}
    \Delta_\text{sg}(t_0+\Delta t) =k_\text{sg} \Delta t.
\end{equation}
Both the drift rate and the offset of each spring gravimeter were determined by fitting the calibration data to simultaneous atomic measurements. This nightly calibration effectively removed short-term instrumental drift and stabilized the instruments' local reference points, as illustrated in Fig.~(\ref{fig:driftcorrection}a,b). After tidal correction, the atomic gravimeter measurements, Fig.~(\ref{fig:driftcorrection}c), exhibit a gradual increase of approximately \SI{50}{\micro\Gal} over the survey period. Given the coastal location of the survey area, this variation may reflect environmental and regional gravity effects not captured by the available tidal correction models; we do not attempt to independently resolve its origin here. Instead, the continuously operating atomic gravimeter provides a direct measurement of the evolving local gravity reference, allowing this variation to be incorporated into the temporal correction of the spring gravimeters. This is particularly important over the extended survey period: a single calibration referenced only to the beginning of the campaign increasingly departs from the evolving absolute reference, as demonstrated by the growing difference between the single (blue) and daily (green) calibration datasets in Fig.~(\ref{fig:driftcorrection}a,b). This difference translates directly into a systematic temporal discrepancies in the field measurements, amounting to approximately $5$-$10~\si{\micro\Gal}$ for measurements on 6 October and increasing progressively to $\sim 50~\si{\micro\Gal}$ for measurements near the end of the survey. Daily calibration therefore prevents temporal variations in the absolute reference from being incorrectly interpreted as instrumental drift or, conversely, from being omitted from the correction, maintaining consistency with the evolving local gravity reference throughout the campaign.

% Both the drift rate and the offset of each spring gravimeter were determined by fitting the calibration data to simultaneous atomic measurements. This nightly calibration effectively removed short-term drift and stabilized the instruments’ local reference points, as illustrated in Fig.~(\ref{fig:driftcorrection}a,b). After tidal correction, the atomic gravimeter measurements, Fig.~(\ref{fig:driftcorrection}c), exhibit a gradual increase of approximately \SI{50}{\micro\Gal} over the survey period. This slow trend likely reflects environmental and regional gravity variations beyond solid Earth tides, predominantly due to ocean loading given the survey’s coastal island location \cite{van2017geophysics}. Because the nightly calibration is tied directly to the atomic reference, these long-term variations are naturally incorporated into the spring gravimeter' temporal corrections. By contrast, a single global calibration based solely on the initial reference period fails to capture this evolving baseline, producing a systematic bias that grows over the course of the survey.

Although the atomic gravimeter exhibited high overall stability, a distinct offset was observed between 3 and 4 October 2023, likely due to mechanical settling of the container floor, consistent with previous field observations \cite{Shettell2024}. This shift was excluded from the calibration of the spring gravimeters. Following this initial adjustment, the instrument remained stable for the remainder of the survey. To quantify performance, the Allan deviation of the atomic gravimeter was evaluated under both field and laboratory conditions over similar time spans, Fig.~(\ref{fig:driftcorrection}d). Laboratory data, collected under rigid mounting and tightly controlled environmental conditions, achieved a noise floor of approximately \SI{2}{\micro\Gal}. During the survey, the minimum Allan deviation was \SI{4}{\micro\Gal}, increasing beyond integration times of ~4 hours (15000 s) as flicker noise began to dominate. These results confirm that the atomic gravimeter maintained sufficient long-term stability to support daily in situ calibration, even under variable environmental conditions.

\subsection{Elevation Correction}

% \begin{figure}[t!]
%     \centering
%     \includesvg[width=0.49\textwidth]{ElevError.svg}
% \caption{Field gravity measurements colored by PPK solution quality. Green points correspond to fully converged solutions with $<$10 cm vertical uncertainty. Blue points indicate partially converged solutions that provided usable elevation estimates despite larger or less well-defined uncertainties. Orange points mark locations where the PPK algorithm did not produce a usable solution; these stations were excluded from elevation-dependent analyses.}
% \label{fig:ppk_quality}
% \end{figure}

\begin{figure}[t!]
    \centering
    \includegraphics[width=0.49\textwidth]{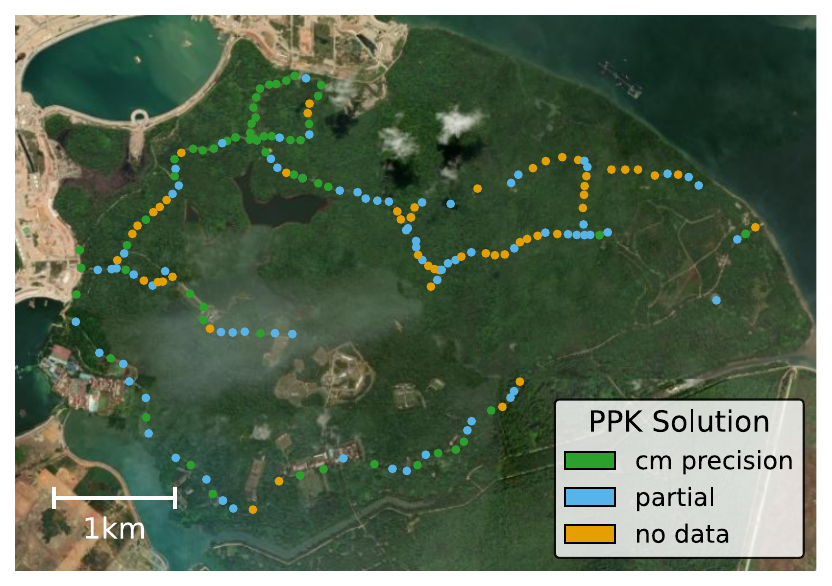}
\caption{Field gravity measurements colored by PPK solution quality. Green points correspond to fully converged solutions with $<$10 cm vertical uncertainty. Blue points indicate partially converged solutions that provided usable elevation estimates despite larger or less well-defined uncertainties. Orange points mark locations where the PPK algorithm did not produce a usable solution; these stations were excluded from elevation-dependent analyses.}
\label{fig:ppk_quality}
\end{figure}

The effect of elevation $h$ at position $\vec{x}$, on the measured gravity signal can be separated into three components \cite{telford1990applied, hinze2013gravity}:
\begin{itemize}
    \item Free-air correction, $g_f(\vec{x})=0.3086\;\text{mGal/m}\times h$, which compensates for the decrease in gravitational acceleration with increasing elevation above the reference level.
    \item Bouguer correction, $g_b(\vec{x}) = 0.04193\;\text{mGal/(m}\cdot\text{g/cm}^{3})\times \rho h$, which restores the gravitational effect of the rock mass between the measurement point and the reference elevation, assuming a uniform subsurface density $\rho=1.6 \text{ g}/\text{cm}^{3}$, representative of compacted tropical topsoil in the surveyed area.
    \item Terrain correction, $g_t(\vec{x})$, which accounts for local topographic deviations from the idealized infinite slab assumed in the Bouguer approximation.
\end{itemize}
The net elevation correction is therefore given by
\begin{equation}
    \label{eq:grav_corr}
    \delta_z g(\vec{x}) = g_f(\vec{x})-g_b(\vec{x})+g_t(\vec{x}).
\end{equation}
In this study $g_t(\vec{x})$ is neglected due to the limited availability of high-resolution topographic data. Given the low relief of the survey area, the resulting terrain correction is estimated to be $\lesssim 1~\si{\micro\Gal}$, well below the measurement precision and dominant sources of uncertainty. Its omission is therefore not expected to affect the reported gravity anomalies or broader spatial trends.

Accurate elevation correction depends on reliable vertical positioning; as uncertainty in the measured elevation propagates directly to the corrected gravity value according to
\begin{equation}
\label{eq:elev_uncertainty}
    \sigma_{g}^{(h)} = \big| \frac{\partial \delta_zg(\vec{x})}{\partial h} \big| \sigma_h \approx 0.242~\si{\milli\Gal}/{\si{\m}} \cdot \sigma_h.
\end{equation}
The distribution of field gravity stations by PPK solution quality is shown in Fig.~(\ref{fig:ppk_quality}). Of the 262 field measurements, 59 yielded fully converged PPK solutions with vertical uncertainties below $\sigma_h=10$ cm, corresponding to an elevation-induced gravity uncertainty of less than $\sigma_{g}^{(h)} \leq  25~\si{\micro\Gal}$; the average elevation-induced uncertainty across these points is approximately $10~\si{\micro\Gal}$. These high-precision points provide the elevation control necessary to validate the drift corrections described in the previous subsection. An additional 127 stations produced partially converged PPK solutions with larger and less well-defined vertical uncertainties; while these measurements carry greater elevation-induced uncertainty, they still capture spatial gravity variations across the survey area and are retained for analysis. Other contributions to the total gravity uncertainty are combined in quadrature and accounted for in the analysis, with elevation uncertainty representing the dominant contribution for the partially converged solutions. Stations for which the PPK solution failed to converge were excluded from the elevation-dependent analysis.

\section{Results}

% \begin{figure}[t!]
%     \centering
%     \includesvg[width=0.49\textwidth]{Results.svg}
% \caption{\emph{Top}: Gravity distribution across the survey after applying temporal and elevation corrections to all stations with a usable PPK solution. \emph{Bottom}: Gravity measurements along the reference path with corresponding high-precision elevation data. Distance 0 corresponds to the westernmost point of the path; upon reaching the junction, the path follows a counterclockwise loop before continuing east. Markers indicate the specific instrument (triangle: spring gravimeter A; square: spring gravimeter B), and colors correspond to the survey day. The profile demonstrates temporal consistency across repeated measurements and reveals a gradual gradient along the path.}
% \label{fig:gravity}
% \end{figure}

\begin{figure}[t!]
    \centering
    \includegraphics[width=0.49\textwidth]{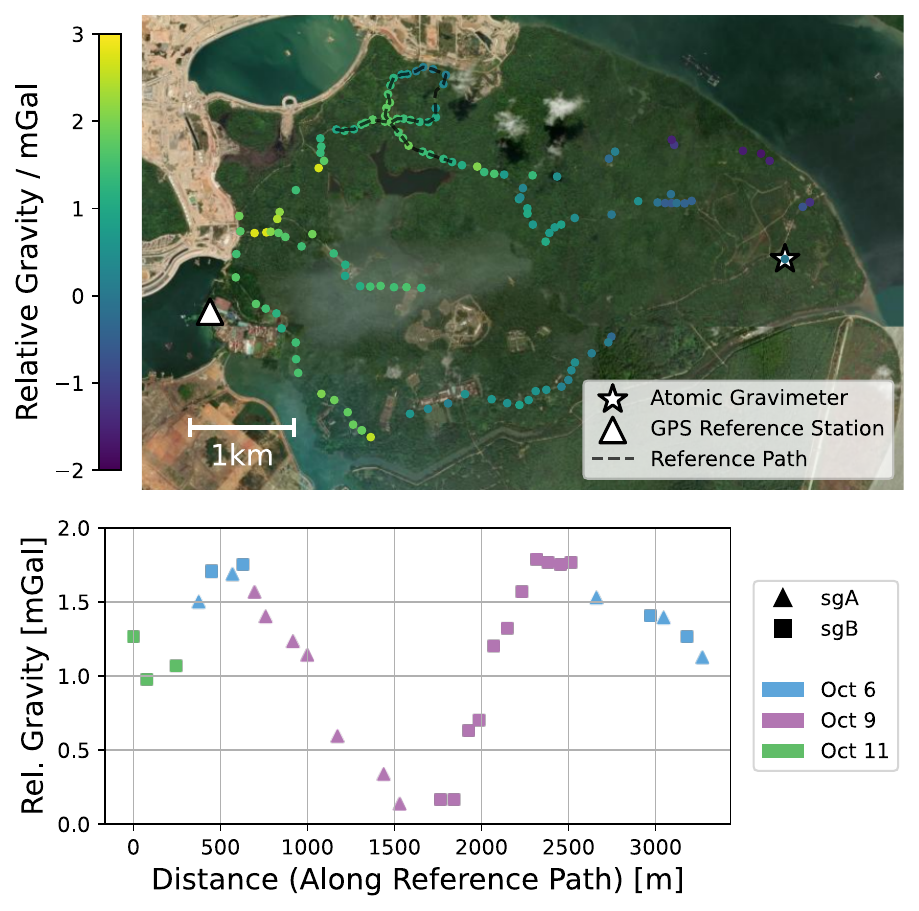}
\caption{\emph{Top}: Gravity distribution across the survey after applying temporal and elevation corrections to all stations with a usable PPK solution. \emph{Bottom}: Gravity measurements along the reference path with corresponding high-precision elevation data. Distance 0 corresponds to the westernmost point of the path; upon reaching the junction, the path follows a counterclockwise loop before continuing east. Markers indicate the specific instrument (triangle: spring gravimeter A; square: spring gravimeter B), and colors correspond to the survey day. Measurements acquired on different survey days recover a consistent gravity profile. The principal uncertainty contributions, summarized in Table~\ref{tab:uncertainty_budget}, combine to a conservatively estimated combined uncertainty of approximately $27~\si{\micro\Gal}$ along the path.}
\label{fig:gravity}
\end{figure}

The final gravity map, corrected for tidal, instrumental, and elevation effects, is shown in Fig.~(\ref{fig:gravity}). Gravity values are referenced to the mean value recorded at the atomic gravimeter site, with the resulting field spanning approximately -2 mGal to +3 mGal across the survey area. A coherent northeast-southwest gradient is observable across the dataset. The survey region contains known geological structures, including two mapped fault lines \cite{dodd2019paleozoic}, providing possible geological context for the observed spatial variations. This large-scale pattern demonstrates that the hybrid survey captured spatially meaningful variations in the gravitational field, even when including stations with reduced vertical precision, and highlights the capability of field-deployed quantum sensors to provide a stable reference for high-precision measurements.

Along a well-sampled reference path within the survey area, many stations yielded fully converged PPK solutions, providing high-precision gravity measurements after drift and elevation correction. These measurements were collected over several days and using both spring gravimeters, enabling robust cross-comparison and validation of the atomic-gravimeter-based drift correction. The resulting gravity profile along this path exhibits a gradual change consistent with the overall trend observed across the survey area. Notably, the looped segment of the path, completed entirely on October 9 with measurements shared between the two spring gravimeters, aligns closely with observations from October 6 at the junction of the reference path, providing additional confirmation of temporal stability. Small residual deviations remain, likely arising from unmodeled instrumental or environmental effects, but the overall pattern demonstrates that the methodology can reliably capture multi-day spatial variations using field-deployed quantum-enabled gravimeters.

Residual deviations between neighboring stations, typically on the order of 100–500 \si{\micro\Gal}, are primarily associated with partially converged PPK solutions, which provide less certain elevation data and therefore introduce uncertainty into the gravity corrections. For example, a vertical positioning uncertainty of 50~cm corresponds to an elevation-induced gravity uncertainty of approximately 125~\si{\micro\Gal} using Eq.~\eqref{eq:elev_uncertainty}. Additional site-dependent effects—including minor soil compaction during measurement, local variations in near-surface density not captured by the uniform Bouguer correction, and environmental influences during deployment—also contribute to the observed scatter. A small fraction of these secondary effects impacted some of the high-precision GPS-supported measurements, introducing additional deviations not explicitly accounted for in the correction model. Importantly, these uncertainties primarily affect the point-to-point spatial accuracy of the partially converged stations, while the measurements remain sufficient to recover the broader regional gravity trends.

% The final gravity map, corrected for both Bouguer effects and instrument drift, is shown in Fig.~(\ref{fig:gravity}). After applying an elevation uncertainty cutoff, the resulting dataset achieves an estimated precision of approximately 0.07mGal, which is sufficient to resolve subsurface density variations on the order of 0.2–0.4 g/cm$^3$ \cite{telford1990applied, 
% hinze2013gravity}. The total gravity variation observed across the survey area spans from roughly 2.5-6mGal, consistent with natural density differences in upper crustal materials \cite{telford1990applied, kearey2013introduction}. While terrain corrections were not included in this study, the level of precision achieved through hybrid gravimetry: a combination of daily atomic gravity reference measurements, mobile spring gravimeters, and filtered GPS data, demonstrates the feasibility of high-resolution gravity surveys in logistically and environmentally challenging environments. This constitutes one of the first large-area field deployments of an atomic gravimeter as a continuously operating calibration reference, highlighting its potential as a robust and portable tool for geophysical field campaigns.

% The full gravitational map after Bouguer and daily drift corrections, accurate to 0.07 mGal based on our elevation uncertainty cutoff, is shown in Fig. \ref{fig:gravity}. The full range of gravitational variations spans between 2.5 to 6 mGal.

\section{Discussion}
This survey represents a significant step toward practical field deployment of hybrid quantum gravimetry. Building on our previous proof-of-concept demonstration under controlled urban conditions, Ref.~\cite{Shettell2024}, we show that continuous quantum calibration remains effective during an extended survey conducted in a realistic tropical environment. By combining a containerized atomic gravimeter with mobile spring instruments, absolute, $\si{\micro\Gal}$ accuracy was propagated across a 24 km$^2$ survey through routine on-site calibration. This enabled asynchronous cross-comparisons between independent gravimeters while suppressing instrumental drift over multiple days. The atomic gravimeter maintained high stability throughout the campaign, including after minor initial mechanical settling of the container floor, confirming its suitability as a reliable field reference. Housed entirely within a single air-conditioned shipping container, the system preserved laboratory-class sensitivity for over two weeks (Fig.~\ref{fig:driftcorrection}d) despite tropical climate conditions and environmental noise, demonstrating robust operation outside the laboratory.

The present implementation imposes practical constraints on survey logistics, primarily through the need for periodic recalibration of the mobile spring gravimeters against the atomic reference and the quality of available positioning data. The required calibration interval depends on the drift characteristics of the relative instruments and the desired gravity accuracy: as the interval between reference measurements increases, departures from the predominantly linear drift become increasingly important and increase the uncertainty of the drift correction. Larger or more distributed surveys may therefore require multiple calibration sites or more frequent reference measurements to maintain the required accuracy. Similarly, under the dense canopy encountered in this campaign, obtaining fully converged PPK solutions required longer observation times, which can increase the duration of individual survey stations. Scaling such surveys may therefore require either longer occupation times or a larger number of mobile spring gravimeters to maintain practical spatial coverage. Thus, while the present campaign demonstrates the approach over 24~km$^2$ and seven days, these dimensions represent the operational configuration of the experiment rather than fundamental limits on the survey area or duration.

The hybrid approach enabled mobile spring gravimeters to resolve gravity variations over a large spatial area and extended timescale. Fully converged PPK stations provided high-precision control, while partially converged stations still contributed to recovering broader gravity trends despite degraded GPS elevation accuracy under dense canopy. These results demonstrate that transportable atomic gravimeters can provide practical, traceable calibration references for realistic geophysical surveys, extending hybrid quantum gravimetry beyond controlled demonstrations. Future platforms will further reduce mobilisation time through autonomous tilt control, streamlined deployment, and cloud-based data processing, broadening access to quantum-enabled gravimetry for applications including hydrology, natural hazard monitoring, resource exploration, and environmental change detection.

\bibliography{main}

@article{kasevich1991atomic,
  title={Atomic interferometry using stimulated Raman transitions},
  author={Kasevich, Mark and Chu, Steven},
  journal={Physical review letters},
  volume={67},
  number={2},
  pages={181},
  year={1991},
  publisher={APS}
}

@article{peters1999measurement,
  title={Measurement of gravitational acceleration by dropping atoms},
  author={Peters, Achim and Chung, Keng Yeow and Chu, Steven},
  journal={Nature},
  volume={400},
  number={6747},
  pages={849--852},
  year={1999},
  publisher={Nature Publishing Group UK London}
}

@article{gillot2014stability,
  title={Stability comparison of two absolute gravimeters: optical versus atomic interferometers},
  author={Gillot, Pierre and Francis, Olivier and Landragin, Arnaud and Dos Santos, F Pereira and Merlet, S{\'e}bastien},
  journal={Metrologia},
  volume={51},
  number={5},
  pages={L15},
  year={2014},
  publisher={IOP Publishing}
}

@article{menoret2018gravity,
  title={Gravity measurements below 10- 9 g with a transportable absolute quantum gravimeter},
  author={M{\'e}noret, Vincent and Vermeulen, Pierre and Le Moigne, Nicolas and Bonvalot, Sylvain and Bouyer, Philippe and Landragin, Arnaud and Desruelle, Bruno},
  journal={Scientific reports},
  volume={8},
  number={1},
  pages={12300},
  year={2018},
  publisher={Nature Publishing Group UK London}
}

@inproceedings{hugill1990scintrex,
  author    = {Andrew Hugill},
  title     = {Scintrex CG-3 automated gravity meter: Description and field results},
  booktitle = {SEG Technical Program Expanded Abstracts 1990},
  pages     = {601--604},
  year      = {1990},
  publisher = {Society of Exploration Geophysicists}
}

@article{gabalda2003cg3tool,
  title={CG3TOOL: an interactive computer program to process Scintrex CG-3/3M gravity data for high-resolution applications},
  author={Gabalda, Germinal and Bonvalot, Sylvain and Hipkin, R},
  journal={Computers \& Geosciences},
  volume={29},
  number={2},
  pages={155--171},
  year={2003},
  publisher={Elsevier}
}

@article{lacoste1988zero,
  title={The zero-length spring gravity meter},
  author={LaCoste, Lucien},
  journal={The Leading Edge},
  volume={7},
  number={7},
  pages={20--21},
  year={1988},
  publisher={Society of Exploration Geophysicists}
}

@article{timmen2012observing,
  title={Observing gravity change in the Fennoscandian uplift area with the Hanover absolute gravimeter},
  author={Timmen, Ludger and Gitlein, Olga and Klemann, Volker and Wolf, Detlef},
  journal={Pure and applied geophysics},
  volume={169},
  pages={1331--1342},
  year={2012},
  publisher={Springer}
}

@article{francis1998calibration,
  title={Calibration of a superconducting gravimeter by comparison with an absolute gravimeter FG5 in Boulder},
  author={Francis, Olivier and Niebauer, TM and Sasagawa, G and Klopping, F and Gschwind, J},
  journal={Geophysical Research Letters},
  volume={25},
  number={7},
  pages={1075--1078},
  year={1998},
  publisher={Wiley Online Library}
}

@article{merlet2021calibration,
  title={Calibration of a superconducting gravimeter with an absolute atom gravimeter},
  author={Merlet, S{\'e}bastien and Gillot, Pierre and Cheng, Bing and Karcher, Romain and Imanaliev, Almazbek and Timmen, Ludger and Pereira dos Santos, Franck},
  journal={Journal of Geodesy},
  volume={95},
  number={5},
  pages={62},
  year={2021},
  publisher={Springer}
}

@incollection{Niebauer2015,
  author    = {Timothy M. Niebauer},
  title     = {Gravimetric Methods -- Absolute and Relative Gravity Meter: Instruments Concepts and Implementation},
  booktitle = {Treatise on Geophysics},
  editor    = {Gerald Schubert},
  edition   = {2nd},
  volume    = {3},
  pages     = {37--57},
  publisher = {Elsevier},
  address   = {Amsterdam},
  year      = {2015},
  isbn      = {978-0-444-53802-4}
}

@article{van2017geophysics,
  title={Geophysics from terrestrial time-variable gravity measurements},
  author={Van Camp, Michel and de Viron, Olivier and Watlet, Arnaud and Meurers, Bruno and Francis, Olivier and Caudron, Corentin},
  journal={Reviews of Geophysics},
  volume={55},
  number={4},
  pages={938--992},
  year={2017},
  publisher={Wiley Online Library}
}

@incollection{Schilling2015accuracy,
  author    = {Manuel Schilling and Olga Gitlein},
  title     = {Accuracy Estimation of the IfE Gravimeters Micro-g LaCoste gPhone-98 and ZLS Burris Gravity Meter B-64},
  booktitle = {IAG 150 Years},
  series    = {International Association of Geodesy Symposia},
  volume    = {143},
  pages     = {249--256},
  publisher = {Springer, Cham},
  year      = {2015}
}

@article{niebauer2011simultaneous,
  title={Simultaneous gravity and gradient measurements from a recoil-compensated absolute gravimeter},
  author={Niebauer, TM and Billson, Ryan and Ellis, Brian and Mason, Bryon and van Westrum, Derek and Klopping, Fred},
  journal={Metrologia},
  volume={48},
  number={3},
  pages={154},
  year={2011},
  publisher={IOP Publishing}
}

@article{greco2012combining,
  title={Combining relative and absolute gravity measurements to enhance volcano monitoring},
  author={Greco, F and Currenti, G and D’Agostino, Giancarlo and Germak, ALESSANDRO and Napoli, R and Pistorio, A and Del Negro, C},
  journal={Bulletin of volcanology},
  volume={74},
  pages={1745--1756},
  year={2012},
  publisher={Springer}
}

@article{chen2025joint,
  title={Joint gravity survey using an absolute atom gravimeter and relative gravimeters},
  author={Chen-yang, Li and Ru-gang, Xu and Xi, Chen and Hong-bo, Sun and Su-peng, Li and Yu, Luo and Ming-qi, Huang and Xue-feng, Di and Zhao-long, Li and Wei-peng, Xiao and others},
  journal={arXiv preprint arXiv:2504.00588},
  year={2025}
}

@article{zerbini2002multi,
  title={Multi-parameter continuous observations to detect ground deformation and to study environmental variability impacts},
  author={Zerbini, S and Negusini, M and Romagnoli, C and Domenichini, F and Richter, B and Simon, D},
  journal={Global and Planetary Change},
  volume={34},
  number={1-2},
  pages={37--58},
  year={2002},
  publisher={Elsevier}
}

@inproceedings{pillai2022potential,
  title={Potential of GNSS Post Processing Kinematic (PPK) technique for test range surveying applications using compact, low cost GNSS Modules},
  author={Pillai, Aiswarya S and Mahato, Somnath and Goswami, Mrinal and Banerjee, P and Bose, Anindya},
  booktitle={2022 URSI Regional Conference on Radio Science (USRI-RCRS)},
  pages={1--4},
  year={2022},
  organization={IEEE}
}

@article{dodd2019paleozoic,
  title={Paleozoic to Cenozoic sedimentary bedrock geology and lithostratigraphy of Singapore},
  author={Dodd, Thomas JH and Gillespie, Martin R and Leslie, A Graham and Kearsey, Timothy I and Kendall, Rhian S and Bide, Thomas P and Dobbs, Marcus R and Millar, Ian L and Lee, Michael Kim Woon and Chiam, Kiefer and others},
  journal={Journal of Asian Earth Sciences},
  volume={180},
  pages={103878},
  year={2019},
  publisher={Elsevier}
}

@article{chapin1996theory,
  title={The theory of the Bouguer gravity anomaly: A tutorial},
  author={Chapin, David A},
  journal={The Leading Edge},
  volume={15},
  number={5},
  pages={361--363},
  year={1996},
  publisher={Society of Exploration Geophysicists}
}

@article{mikolaj2013first,
  title={The first tidal analysis based on the CG-5 Autograv gravity measurements at Modra station},
  author={Mikolaj, Michal and H{\'a}bel, Branislav},
  journal={Contributions to Geophysics and Geodesy},
  volume={43},
  number={1},
  pages={59--72},
  year={2013}
}

@article{jentzsch2005earth,
  title={Earth tides and ocean tidal loading},
  author={Jentzsch, Gerhard},
  journal={Tidal phenomena},
  pages={145--171},
  year={2005},
  publisher={Springer}
}

@article{davis2008time,
  title={Time-lapse gravity monitoring: A systematic 4D approach with application to aquifer storage and recovery},
  author={Davis, Kristofer and Li, Yaoguo and Batzle, Michael},
  journal={Geophysics},
  volume={73},
  number={6},
  pages={WA61--WA69},
  year={2008},
  publisher={Society of Exploration Geophysicists}
}

@inproceedings{nishijima2016repeat,
  title={Repeat absolute and relative gravity measurements for geothermal reservoir monitoring in the Ogiri Geothermal Field, Southern Kyushu, Japan},
  author={Nishijima, J and Umeda, C and Fujimitsu, Y and Takayama, J and Hiraga, N and Higuchi, S},
  booktitle={IOP Conference Series: Earth and Environmental Science},
  volume={42},
  number={1},
  pages={012004},
  year={2016},
  organization={IOP Publishing}
}

@article{sugihara2008geothermal,
  title={Geothermal reservoir monitoring with a combination of absolute and relative gravimetry},
  author={Sugihara, Mituhiko and Ishido, Tsuneo},
  journal={Geophysics},
  volume={73},
  number={6},
  pages={WA37--WA47},
  year={2008},
  publisher={Society of Exploration Geophysicists}
}

@article{ruan2024transportable,
  title={A transportable atomic gravimeter with constraint-structured active vibration isolation},
  author={Ruan, Chuanjing and Zhuang, Wei and Yao, Jiamin and Zhao, Yang and Ma, Zenghan and Yi, Cong and Tian, Qin and Wu, Shuqing and Fang, Fang and Wen, Yinghong},
  journal={Sensors},
  volume={24},
  number={8},
  pages={2395},
  year={2024},
  publisher={MDPI}
}

@article{Shettell2024,
  title = {Geophysical survey based on hybrid gravimetry using relative measurements and an atomic gravimeter as an absolute reference},
  volume = {14},
  ISSN = {2045-2322},
  url = {http://dx.doi.org/10.1038/s41598-024-57253-1},
  number = {1},
  journal = {Scientific Reports},
  publisher = {Springer Science and Business Media LLC},
  author = {Shettell,  Nathan and Lee,  Kai Sheng and Oon,  Fong En and Maksimova,  Elizaveta and Hufnagel,  Christoph and Wei,  Shengji and Dumke,  Rainer},
  year = {2024},
  month = mar 
}

@book{telford1990applied,
  author    = {Telford, William Murray and Geldart, Lloyd P.
               and Sheriff, Robert E.},
  title     = {Applied Geophysics},
  edition   = {2nd},
  publisher = {Cambridge University Press},
  address   = {Cambridge},
  year      = {1990}
}

@book{hinze2013gravity,
  author    = {William J. Hinze and Ralph R. B. von Frese and Afif H. Saad},
  title     = {Gravity and Magnetic Exploration: Principles, Practices, and Applications},
  publisher = {Cambridge University Press},
  year      = {2013},
  address   = {Cambridge},
  isbn      = {978-0521871013}
}

@article{flechtner2021satellite,
  title={Satellite gravimetry: a review of its realization},
  author={Flechtner, Frank and Reigber, Christoph and Rummel, Reiner and Balmino, Georges},
  journal={Surveys in Geophysics},
  volume={42},
  number={5},
  pages={1029--1074},
  year={2021},
  publisher={Springer}
}

@article{imanishi2002calibration,
  title={Calibration of the superconducting gravimeter T011 by parallel observation with the absolute gravimeter FG5\# 210—a Bayesian approach},
  author={Imanishi, Yuichi and Higashi, Toshihiro and Fukuda, Yoichi},
  journal={Geophysical Journal International},
  volume={151},
  number={3},
  pages={867--878},
  year={2002},
  publisher={Blackwell Publishing Ltd Oxford, UK}
}

\vspace{5pt}

\noindent\emph{Funding -} This research is supported by the NRF through NRF2021-QEP2-03-P06.

\vspace{5pt}

\noindent\emph{Acknowledgements -} The authors would like to thank Johnathan Kim, for organizing the logistics of the geophysical survey, Aveek Chandra, Gan Koon Siang, Christoph Hufnagel and Paul Tan for assisting with the gravity measurements as well as, Maung Maung Phyo, Yukuan Chen, Yu Jiang, Nurdin Elon Dahlan, and Shengji Wei for assisting with preparations of the geophysical survey, and Jan Mrlina for the insightful discussions on geology.

\vspace{5pt}

\noindent\emph{Author Contributions -} N.S. and K.S.L. conducted the data analysis. All authors conducted the data collection during the survey. F.E.O. operated the atomic gravimeter. R.D. conceived the project. N.S., K.S.L., and R.D. contributed to the writing of the manuscript.

\vspace{5pt}

\noindent\emph{Data Availability -} Raw data captured from the relative gravimeter, absolute gravimeter, and gps during the survey, can be provided by N.S. upon reasonable request.

\vspace{5pt}

\noindent\emph{Competing Interests -} The authors have no conflicts to disclose.

%\clearpage
%\widetext
%\appendix 

\end{document}